\documentclass[conference]{IEEEtran}
\usepackage{amsmath,graphicx}
\usepackage{amsfonts,amssymb,dsfont}
\usepackage{multirow}
\usepackage{algorithm}
\usepackage{algpseudocode}
\usepackage{cite}
\usepackage{balance}
\DeclareGraphicsExtensions{.jpg,pdf,.png}
\ifCLASSINFOpdf
\else
\fi
\usepackage{fancyhdr}
\newcommand*{\myfont}{\fontfamily{phv}\selectfont}
\DeclareTextFontCommand{\textmyfont}{\myfont}

\begin{document}

\title{Bounds on Discrete Fourier Transform of\\Random Mask}
\author{Nematollah Zarmehi and Farokh Marvasti\\Advanced Communication Research Institute\\Department of Electrical Engineering\\Sharif University of Technology, Tehran, Iran\\Email: http://zarmehi.ir/contact.html, marvasti@sharif.edu}

\maketitle
\thispagestyle{fancy}
\begin{abstract}
This paper proposes some bounds on the maximum of magnitude of a random mask in Fourier domain. The random mask is used in random sampling scheme. Having a bound on the maximum value of a random mask in Fourier domain is very useful for some iterative recovery methods that use thresholding operator. In this paper, we propose some different bounds and compare them with the empirical examples.
\end{abstract}

\section{Introduction}
There are different sampling schemes in the field of digital signal processing. The primary sampling theorem known as Nyquist-Shannon theorem states that a band-limited signal can be recovered from its uniform samples under certain condition \cite{ref:shannon,ref:landau,ref:marvastibook}. The uniform sampling was used for some decades. Later, other sampling methods such as non-uniform sampling \cite{non1,non2}, periodic non-uniform sampling \cite{pnon1,pnon2}, and random sampling \cite{rand1,rand2} were proposed.

\begin{figure}[h!]
	\centering
	\begin{minipage}[b]{1\linewidth}
		\centerline{\includegraphics[width=0.66\linewidth]{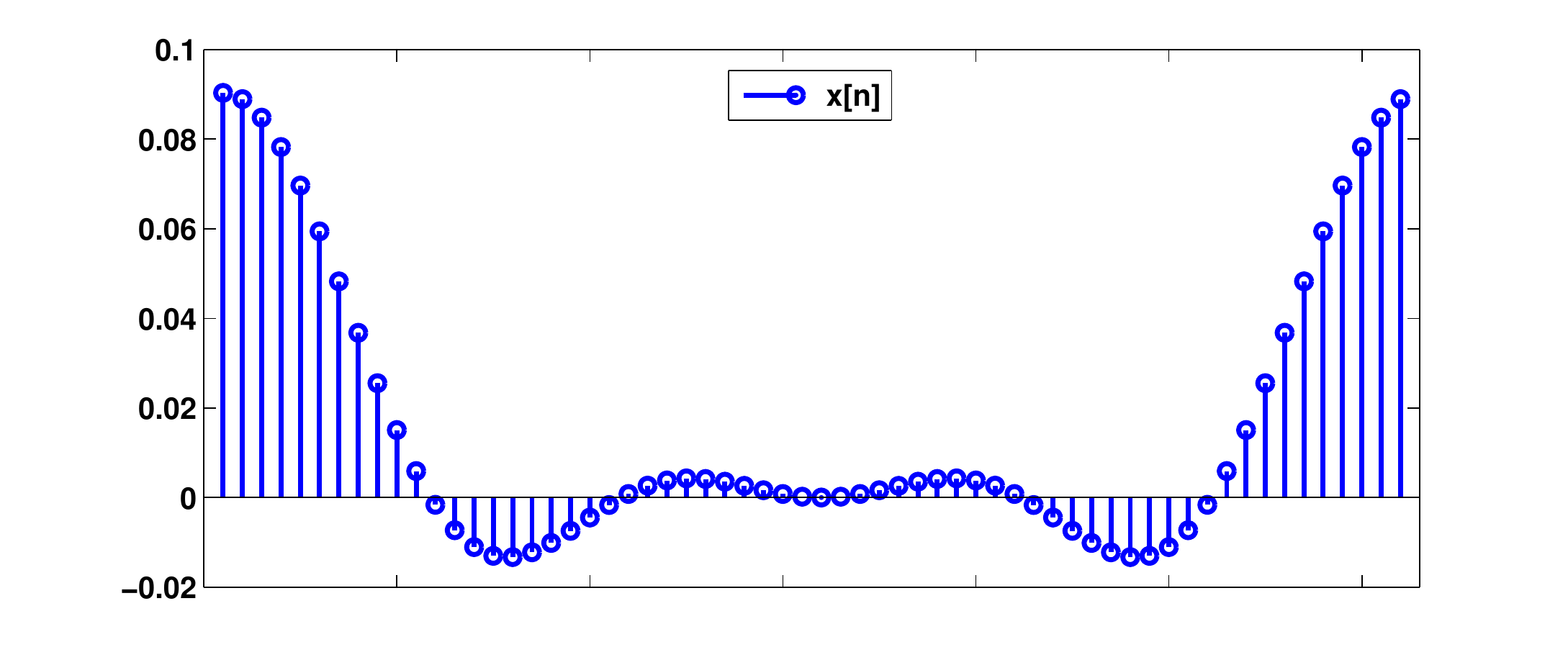}}
		\centerline{a) $x[n]$}\medskip
	\end{minipage}\\ \vspace{0.029cm}
	\begin{minipage}[b]{1\linewidth}
		\centerline{\includegraphics[width=0.66\linewidth]{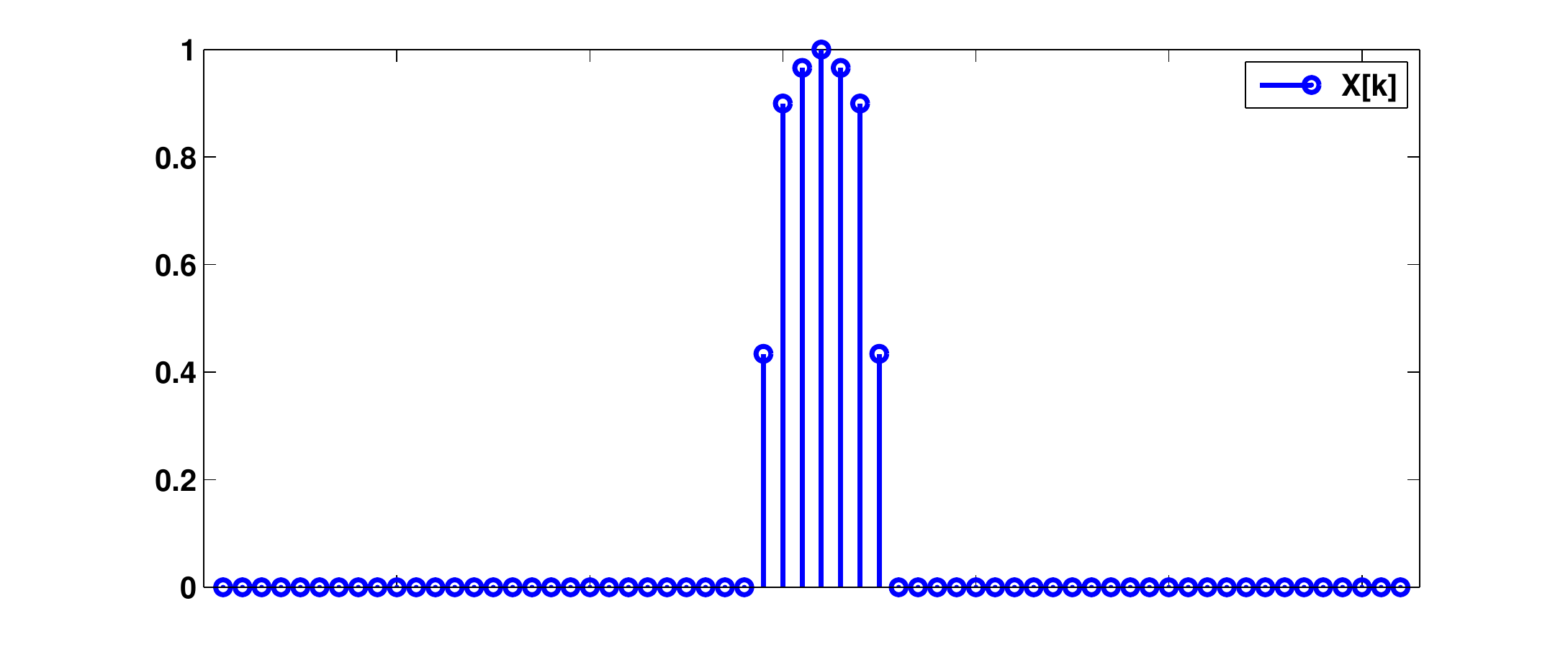}}
		\centerline{b) $|X[k]|$}\medskip
	\end{minipage}
	\caption{An example of band-limited signal: a) $x[n]$ in time domain and b) $X[k]$ in frequency domain.}\label{fig:x}
\end{figure}

\begin{figure}[h!]
	\centering
	\begin{minipage}[b]{1\linewidth}
		\centerline{\includegraphics[width=0.66\linewidth]{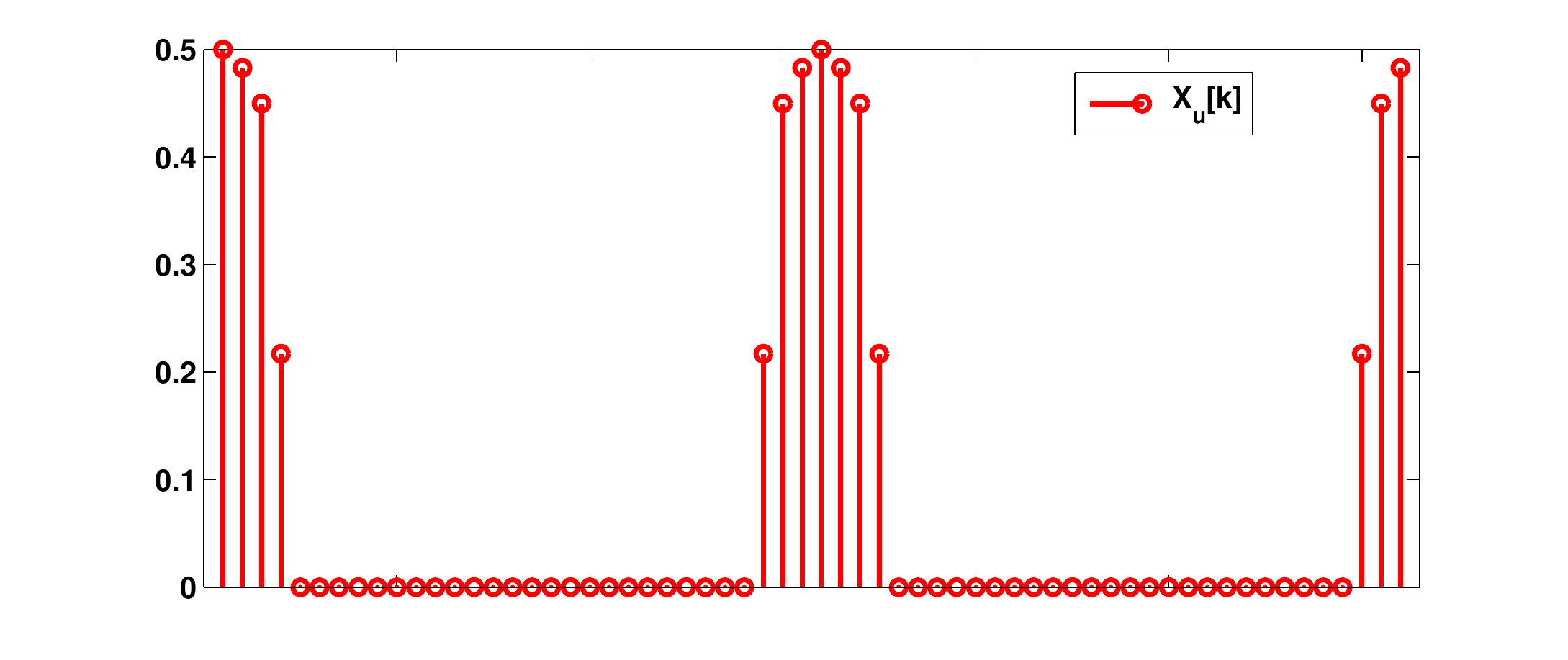}}
		\centerline{a) $|X_u[k]|$}\medskip
	\end{minipage}\\ \vspace{0.029cm}
	\begin{minipage}[b]{1\linewidth}
		\centerline{\includegraphics[width=0.66\linewidth]{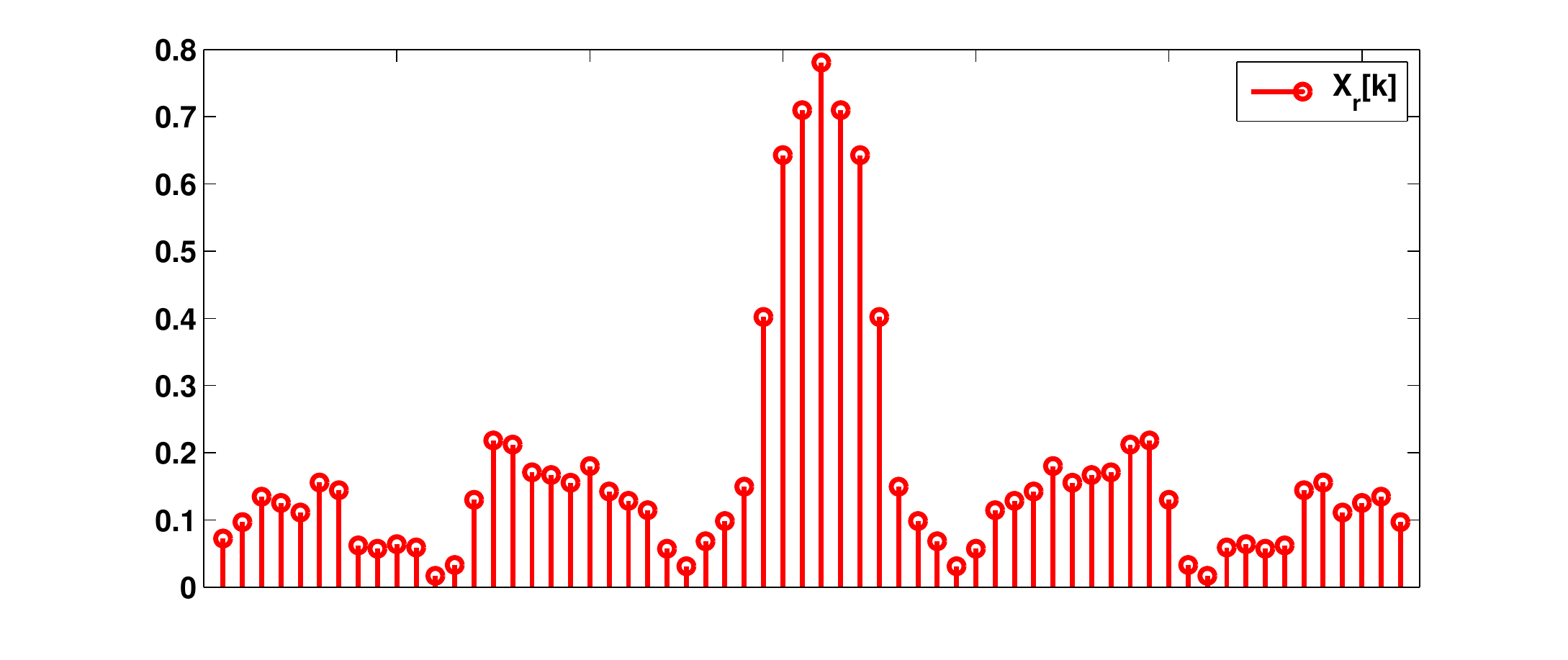}}
		\centerline{b) $|X_r[k]|$}\medskip
	\end{minipage}
	\caption{Fourier transform of uniform and random sampled versions of $x[n]$ that is shown in Fig. \ref{fig:x}}\label{fig:xx}
\end{figure}

In random sampling, the sampler picks up the samples of the original signal at random. It is like that the original signal is multiplied by a random mask containing i.i.d Bernoulli random variables with parameter $p$. Here, $p$ is the sampling rate. There is a major difference between random and uniform sampling in the frequency domain. In uniform sampling scheme, we see that the spectrum of the original signal is repeated by a period equal to the sampling frequency while in random sampling scheme, we see a spectrum similar to the original that does not have periodicity. Let $x[n]$ be the original signal and $x_r[n]$ and $x_u[n]$ be the random sampled and uniform sampled versions of $x[n]$ with sampling rate $0.5$. Fig. \ref{fig:x} shows an example of discrete signal and its Fourier transform and Fig. \ref{fig:xx} shows the magnitude of Fourier transform of $x_u[n]$ and $x_r[n]$. It can be seen that in random sampling scheme, we have no periodicity. Moreover, $X_r[k]$ is very similar to $X[k]$ and there are aliasing components that are produced due to convolving the original signal by the random mask. 

In this paper, we aim to bound the magnitude of Fourier transform of a random mask. The results are very useful for some iterative recovery methods such as IMAT (Iterative Method with Adaptive Thresholding \cite{ref:imat,ref:imatsite}) that use a thresholding operator. We propose some bounds on the magnitude of Fourier transform of a random mask in Section \ref{sec:prp}. In Section \ref{sec:num}, some numerical examples are presented and Finally Section \ref{sec:conc} concludes the paper.

\section{Proposed Bounds}\label{sec:prp}
Suppose that $\mathbf{a}$ is a sequence of $N$ i.i.d Bernoulli random variables with parameter $p$. We aim to find an upper bound for absolute value of $\{\mathbf{A}\}_{k=0}^{N-1}$ as the DFT of $\mathbf{a}$:
\begin{equation}
	{\mathbf{A}_k} = \sum\limits_{n = 1}^{N - 1} {{\mathbf{a}_n}{e^{\frac{{j2\pi kn}}{N}}}} \quad \quad {\rm for} \quad \quad k=0,1,...,N-1.
\end{equation}

\subsection{The Worst Case}
We know that the average number of ones in $\mathbf{a}$ is $Np$. Therefore, on average, $Np$ is an upper bound for $A_0$. Define the support set $\mathcal{P}$ such that $\mathbf{a}_i=1$, $\forall i \in \mathcal{P}$. It is clear that $\mathcal{P} \subset \{ 0,1,...,N-1 \}$ and $|\mathcal{P}| \triangleq N_p \approx \left\lceil Np \right\rceil$. Moreover, suppose that $N$ is prime. We have:

\begin{equation}\label{eq:abs1}
	\begin{split}
		|A_k| &= \left|{ \sum\limits_{n = 1}^{N - 1} {{{\bf a}_n}{e^{\frac{{j2\pi kn}}{N}}}} } \right|  = \left|{ \sum\limits_{\theta_{k,n} \in \Theta_k}^{} {{e^{j\theta_{k,n}}}} } \right| \\
		{} &  = \left|{ \sum\limits_{\theta_{k,n} \in \Theta_k}^{} \left[ {\cos(\theta_{k,n})+ j\sin(\theta_{k,n})} \right] } \right| \\
		{}& =\left|{ \sum\limits_{n \in \mathcal{P}}^{} \cos(\theta_{k,n}) + j \sum\limits_{n \in \mathcal{P}}^{} \sin(\theta_{k,n})  } \right| \\
		{} & = \sqrt{     \left({\sum\limits_{n \in \mathcal{P}}^{} \cos(\theta_{k,n}) }\right)^2   + \left( {  \sum\limits_{n \in \mathcal{P}}^{} \sin(\theta_{k,n})  }\right)^2          }\\
		{} & 	=\sqrt{N_p + 2.\sum\limits_{{m,n \in \mathcal{P}} \over m\neq n}^{} \cos(\theta_{k,m}-\theta_{k,n}) },
	\end{split}
\end{equation}
where $\Theta_k=\left\{ {\theta_{k,n}=\frac{2k\pi n}{N} ~~|~~ n \in \mathcal{P} }\right\} $  for all $k \in \mathcal{K}$. The expression in (\ref{eq:abs1}) will be maximized if $\theta_{k,m} ~ {\buildrel 2\pi \over =} ~ \theta_{k,n}$. But this is impossible because:
\begin{equation}\label{eq:n2}
	\nexists~ \theta_{k,m},\theta_{k,n} \in \Theta_k: \quad \quad \theta_{k,m} ~ {\buildrel 2\pi \over \neq} ~ \theta_{k,n}. 
\end{equation}
Hence, the presented expression will be maximized when $\theta_{k,m}$ is very close to $\theta_{k,n}$. The minimum distance between $\theta_{k,m}$ and $\theta_{k,n}$ is equal to $2\pi/N$. In other words, maximum value will be achieved when ${\bf a}$ contains an all-one vector ${\bf u}=[1, 1, ..., 1]$ of length $N_p$.

After some mathematical computations, we can get the maximum value of $\{|A_k|\}_{k=1}^{N-1}$ as follows:
\begin{equation}\label{max}
	|A_k|_{max} =\sqrt{N_p + 2.\sum\limits_{i=1}^{N_p-1} (N_p-i)cos\left({\frac{2\pi}{N}i}\right) }.
\end{equation}

Table \ref{tab:s1} shows the upper bounds for $|A_k|$, $\frac{|A_k|}{N_p}$, and the maximum values obtained by simulation for different values of $N$ and $p$.

\begin{table*}[t!]
\begin{center}
\caption{Comparison between the proposed upper bound and the maximum value achieved by simulation.}\label{tab:s1} 
\renewcommand{\arraystretch}{1.0}
{\small
\begin{tabular}{|c|c|c|c|c|c|c|}
\hline
$N$ & $p$ & $N_p$ & {$|A_k|_{max}$ (Sim.)}   & { ${\frac{|A_k|}{N_p}}_{max}$ (Sim.)} & \textbf{$|A_k|_{max}$}   &  ${\frac{|A_k|}{N_p}}_{max}$  \\ \hline  \hline
127 & 0.5 & 64  &   11.55  &   0.182  &  40.426 & 0.637\\ \hline
127 & 0.8 & 102  &   12  &   0.118  &  23.439 & 0.231\\ \hline
127 & 0.1 & 13  &   7.705  &   0.607  &  12.778 & 1.006\\ \hline
1543 & 0.5 & 772  &   52.383  &   0.068  &  491.152 & 0.637\\ \hline
127 & 0.8 & 1235  &   617.2  &   0.033 &  288.207 & 0.233 \\ \hline
127 & 0.1 & 155  &   38.628  &   0.25  &  152.44 & 0.988 \\ \hline
131071 & 0.5 & 65535  &   618.651  &   0.0094  & 4.172$\times10^4$  & 0.637 \\ \hline
131071 & 0.8 & 104856  &   471.001  &   0.0045 &  2.452$\times10^4$ & 0.234 \\ \hline
131071 & 0.1 & 13107  &   344.254  &   0.026  &  1.289$\times10^4$ & 0.984 \\ \hline
\end{tabular}}
\end{center}
\vspace{-0.4cm}
\end{table*}

According to Table \ref{tab:s1}, the ratio $|A_k|/N_p$ is almost constant. We can approximate $|A_k|/N_p$ as shown in Fig. \ref{fig:appx}.

\begin{figure*}
\begin{equation*}
\begin{split}
\frac{A_k}{N_p} & =\frac{1}{{{N_p}}}\sqrt {{N_p} + 2\sum\limits_{i = 1}^{{N_p} - 1} {i\cos \left( {\left( {{N_p} - i} \right)\frac{{2\pi i}}{N}} \right)} } \\
{} & =\frac{1}{{{N_p}}}\sqrt {{N_p} + 2\left[ {\cos \left( {{N_p}\frac{{2\pi }}{N}} \right)\sum\limits_{i = 1}^{{N_p} - 1} {i\cos \left( {\frac{{2\pi i}}{N}} \right)}  + \sin \left( {{N_p}\frac{{2\pi }}{N}} \right)\sum\limits_{i = 1}^{{N_p} - 1} {i\sin \left( {\frac{{2\pi i}}{N}} \right)} } \right]}
\end{split}
\end{equation*}
{\scriptsize{
\begin{equation*}
\begin{split}
{} & ={\tiny{ \frac{1}{{{N_p}}}\sqrt {{N_p} + 2\left[ {\frac{{{N_p}}}{2}\frac{{\sin \left( {\frac{{{N_p} - 1}}{N}\pi } \right)\cos \left( {\frac{{{N_p}}}{N}\pi } \right)}}{{\sin \left( {\frac{\pi }{N}} \right)}} + \frac{1}{2}\frac{{\sin \left( {\frac{{{N_p} - 1}}{N}\pi } \right)\cos \left( {\frac{{{N_p}}}{N}\pi } \right)\cos \left( {\frac{\pi }{N}} \right)\sin \left( {\frac{{{N_p}}}{N}\pi } \right)}}{{{{\sin }^2}\left( {\frac{\pi }{N}} \right)}} - \left( {\frac{{{N_p} - 1}}{2}} \right)\frac{{\cos \left( {\frac{{{N_p} - 1}}{N}\pi } \right)sin\left( {\frac{{{N_p}}}{N}\pi } \right)}}{{\sin \left( {\frac{\pi }{N}} \right)}}} \right]} }}
\end{split}
\end{equation*}}}

\begin{equation}\label{app}
\hspace{-3.1cm} \approx \frac{1}{{Np}}\sqrt {Np + \frac{{{N^2}}}{{{\pi ^2}}}{{\sin }^2}\left( {p\pi } \right) - N\left[ {\sin \left( {p\pi } \right) - \frac{1}{{2\pi }}\sin \left( {2p\pi } \right)} \right]} 
\end{equation}
\caption{The approximation of $\frac{A_k}{N_p}$.}\label{fig:appx}
\end{figure*}
		
\begin{figure}[h!]
	\centering
	\includegraphics[width=.91\linewidth]{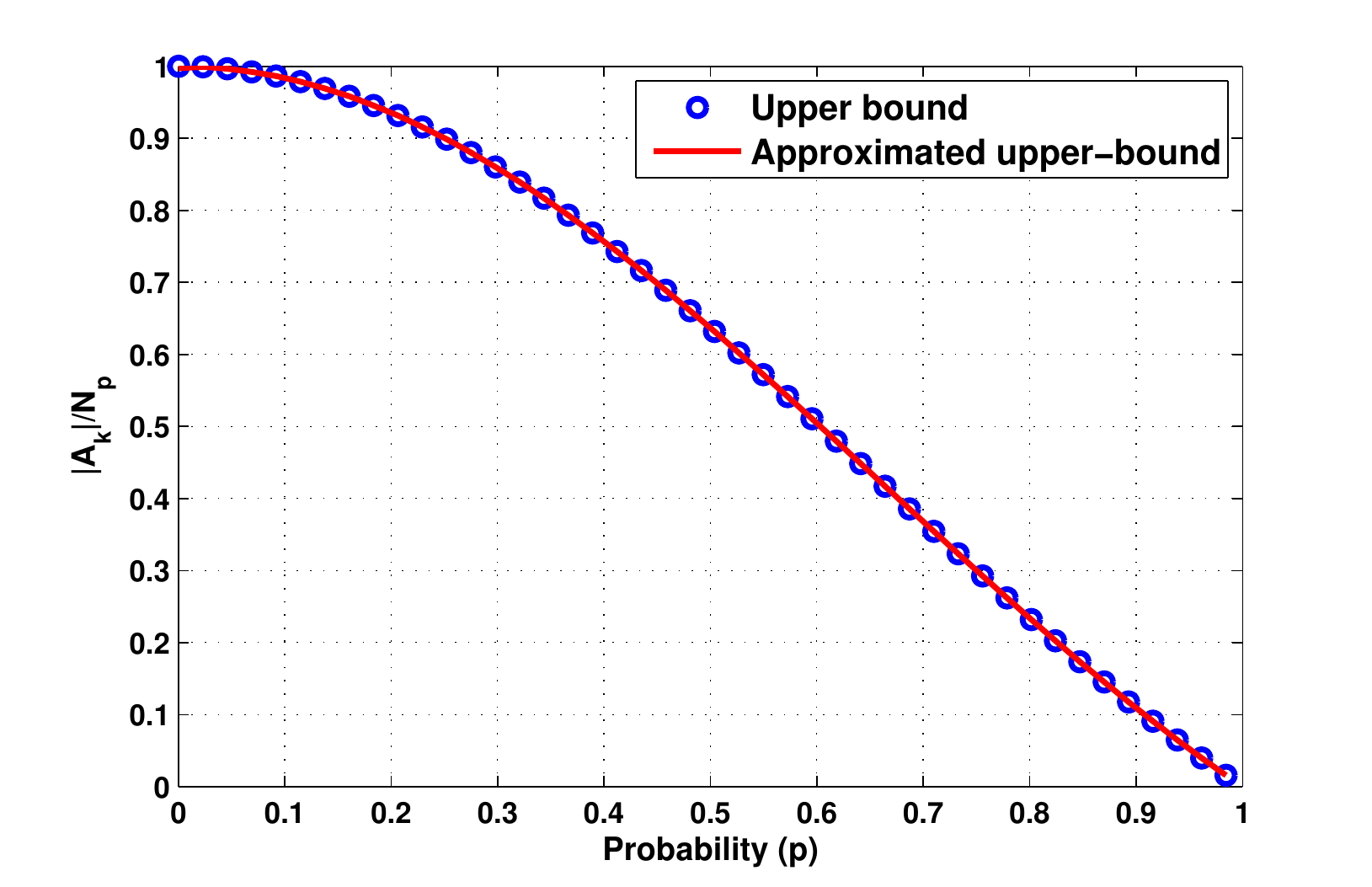}
	\caption{Upper bound and its approximation.}\label{fig:app}
\end{figure}

The upper bound of $\frac{A_k}{N_p}$ and its approximation are depicted in Fig. \ref{fig:app}. It can be seen that the original curve has been followed very well by the approximation curve.

\begin{figure}[h!]
	\centering
	\includegraphics[width=.91\linewidth]{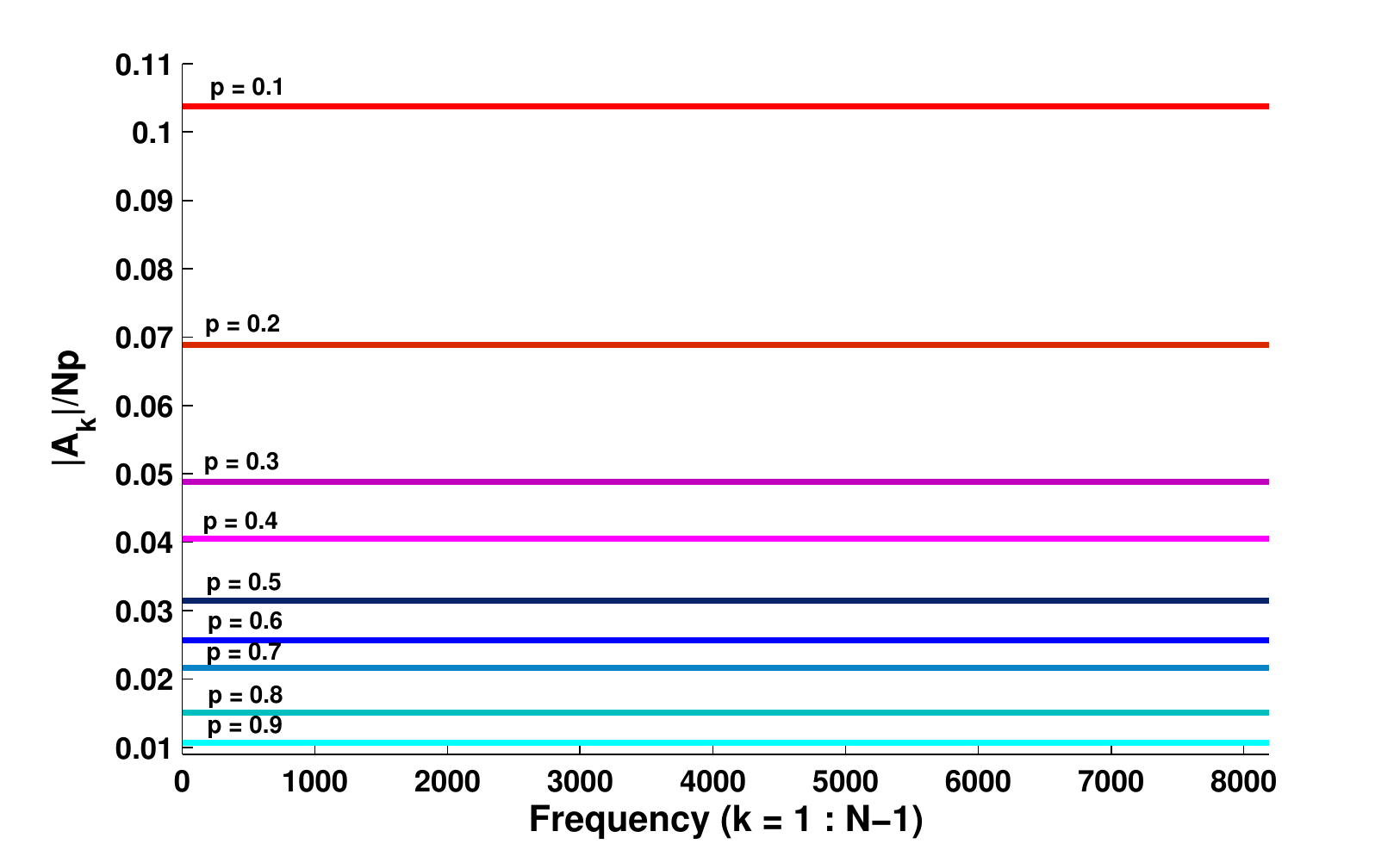}
	\caption{Noise Ratio, $N=2^{13}-1$.}\label{fig:ratio}
\end{figure}

Fig. \ref{fig:ratio} shows the maximum noise ratio $\frac{|A_k|}{Np}$ versus $k$ for different values of $p$. As expected, when the sampling rate is high the noise level is low.

\subsection{Gaussian Approximation}\label{subsec:gauss}
In this Subsection, we approximate a Bernoulli distribution with parameter $p$ by a Gaussian distribution with mean $p$ and variance $p(1-p)$. We want to be sure that \textit{almost all} $\{\mathbf{A}_k\}_{k=1}^{N-1}$ are below than a threshold $T$, i.e., for an arbitrary positive $\epsilon$:
\begin{equation}
	P\bigg({\exists ~k\neq 0~ :~ |\mathbf{A}_k|>T}\bigg) \leq \epsilon.
\end{equation}
From the union bound we have:
\begin{equation}
	\begin{split}
		P\bigg({\exists ~k\neq 0~ :~ |\mathbf{A}_k|>T}\bigg) & \leq P\left({\bigcup_{k=1}^{N-1} \mathds{1}\left[|\mathbf{A}_k|>T\right]}\right)\\
		{} & \leq \sum\limits_{k=1}^{N-1}P\bigg({|\mathbf{A}_k|>T}\bigg),
	\end{split}
\end{equation}
where $\mathds{1}\left[|\mathbf{A}_k|>T\right]$ indicates that $|\mathbf{A}_k|>T$ with probability of 1.

It is not to difficult to verify the fact that the real and imaginary parts of $\mathbf{A}_k$ can be approximated by a Gaussian distribution $\mathcal{N}\left({0,p(1-p)N}\right)$.

Therefore, $P\bigg({|\mathbf{A}_k|>T}\bigg) \leq \epsilon$ is equivalent to $P\bigg({(Re\{\mathbf{A}_k\})^2+(Img\{\mathbf{A}_k\})^2>T^2}\bigg) \leq \epsilon$ and we have:
{\small{
\begin{equation}
\begin{split}
P\bigg({(Re\{\mathbf{A}_k\})^2+(Img\{\mathbf{A}_k\})^2>T^2}\bigg) & \leq P\bigg({(Re\{\mathbf{A}_k\})^2>\frac{T^2}{2}}\bigg) \\ & +P\bigg({(Re\{\mathbf{A}_k\})^2>\frac{T^2}{2}}\bigg).
\end{split}
\end{equation}}}
Hence, we are interested to calculate the following probability:
\begin{equation}
P\left({|Re\{\mathbf{A}_k\}|>\frac{T}{\sqrt{2}}}\right) \leq \frac{\epsilon}{2}
\end{equation}
Since $Re\{\mathbf{A}_k\} \approx \mathcal{N}\left(o,p(1-p)N\right)$, $P\left({|Re\{\mathbf{A}_k\}|>\frac{T}{\sqrt{2}}}\right)=Q\left({\frac{T}{\sqrt{2p(1-p)N}}}\right)$. Hence, we can calculate the maximum of $\{A_k\}_{k=1}^{N-1}$ as follows:
\begin{equation}
Q\left({\frac{T}{\sqrt{2p(1-p)N}}}\right) \leq \frac{\epsilon}{2} ~~\Longrightarrow ~~ T \leq \sqrt{2p(1-p)N}Q^{-1}\left({\frac{\epsilon}{2}}\right)
\end{equation}
One of the most common approximations for Q-function is as follows:
\begin{equation}
Q(x) \approx \frac{1}{2}e^{-\frac{x^2}{2}}.
\end{equation}
If we use above approximation for Q-function, we can approximate $T$ as below:
\begin{equation}
T \leq 2\sqrt{p(p-1)N\log(\epsilon)}
\end{equation}
	
\begin{figure}[h!]
	\centering
	\includegraphics[width=0.9\linewidth]{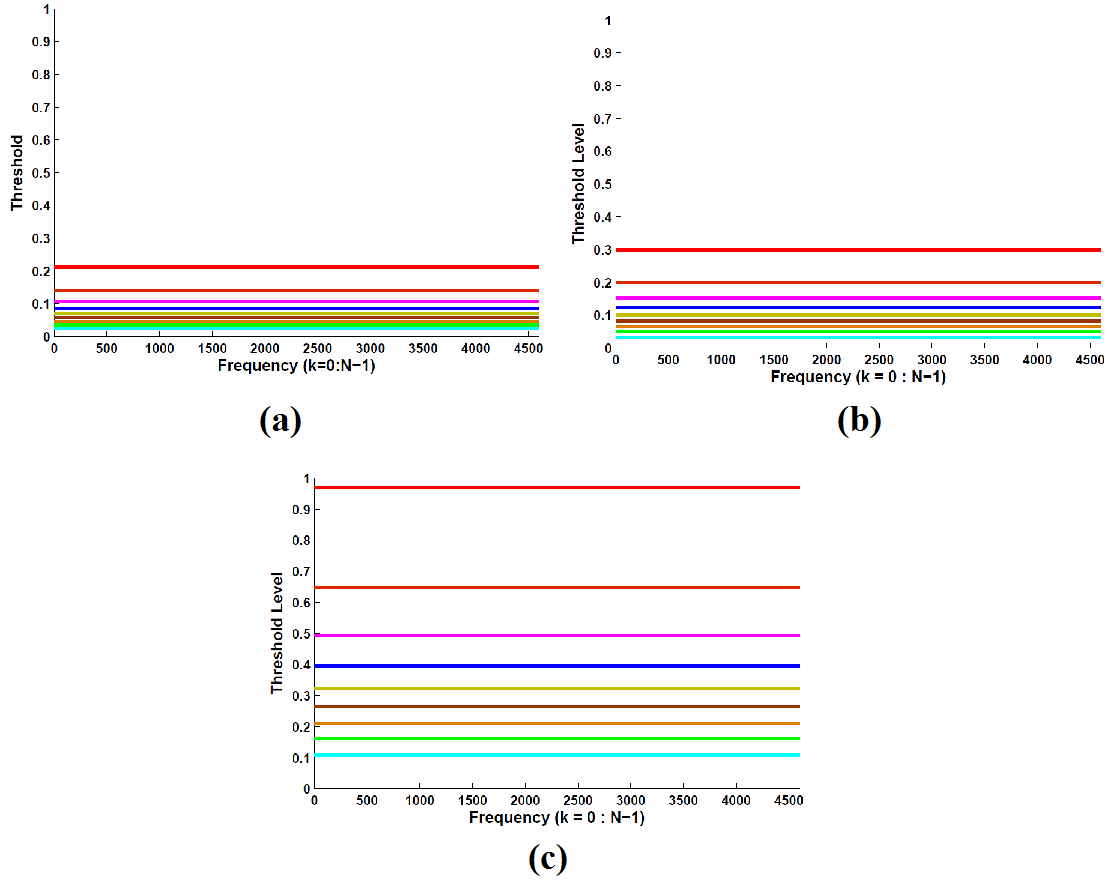}
	\caption{Threshold levels for (a) $\epsilon = 10^{-1}$, (b) $\epsilon = 10^{-6}$, and (c) $\epsilon = 10^{-100}$.}
	\label{fig:thr}
\end{figure}

Fig. \ref{fig:thr} shows the threshold levels for different values of $\epsilon$. It can be seen that as $\epsilon$ is small the threshold levels obtained by the Gaussian approximation is very close to the worst case.

\subsection{Three and Four Sigma Bounds}
From probability and statistical analysis, we know that a sequence of numbers is within about three or four standard deviations from the mean. In Subsection \ref{subsec:gauss}, we approximated the random mask with a Gaussian distribution. Therefore, $3-\sigma$ and $4-\sigma$ bounds for the magnitude of Fourier transform of a random mask are $3\sqrt{p(1-p)N}$ and $4\sqrt{p(1-p)N}$, respectively.

\section{Numerical Examples}\label{sec:num}
\begin{figure}[h!]
	\centering
	\includegraphics[width=.91\linewidth]{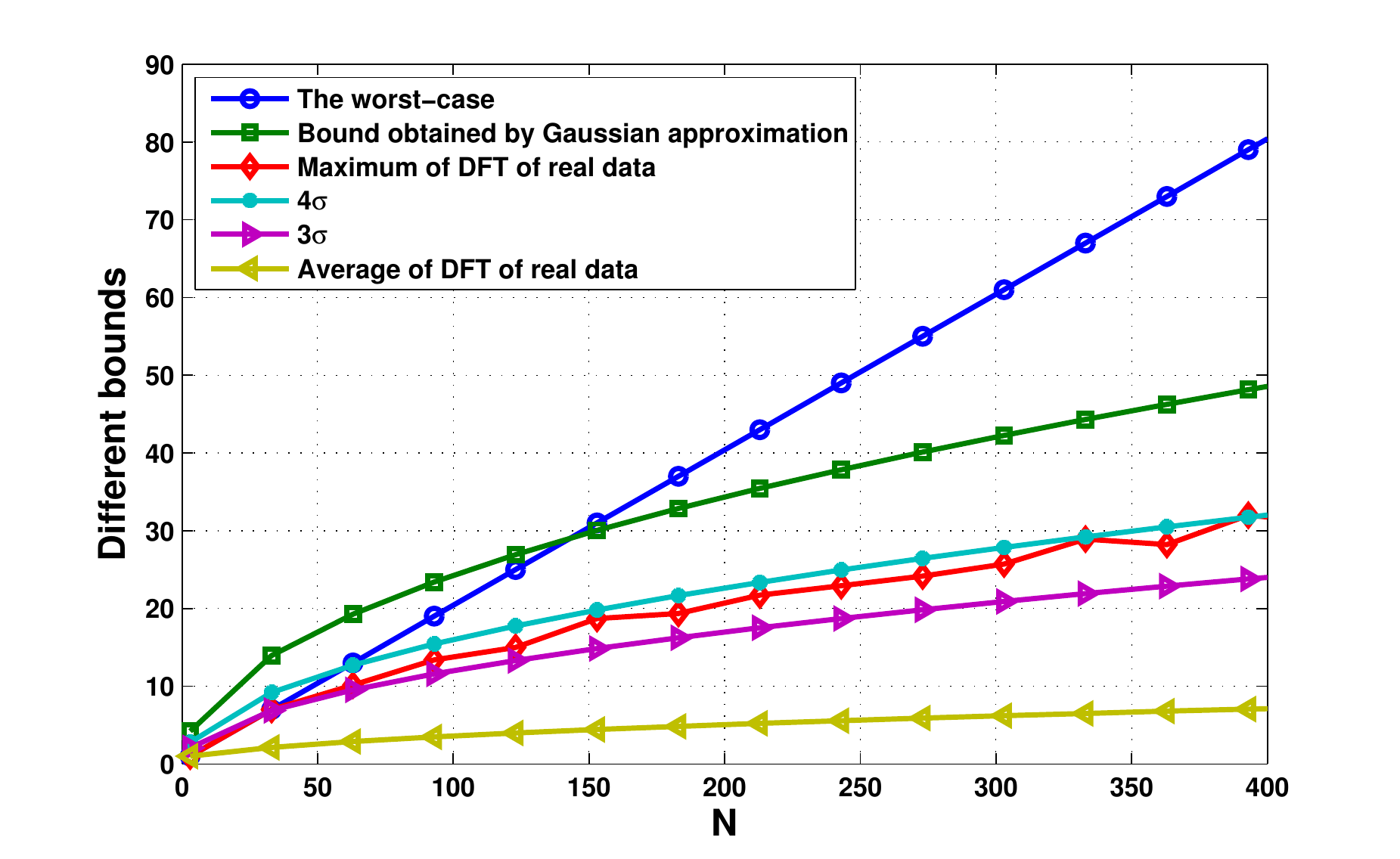}
	\caption{Proposed bounds when the sampling rate is $0.2$.}
	\label{fig2part}
\end{figure}

\begin{figure}[h!]
	\centering
	\includegraphics[width=.91\linewidth]{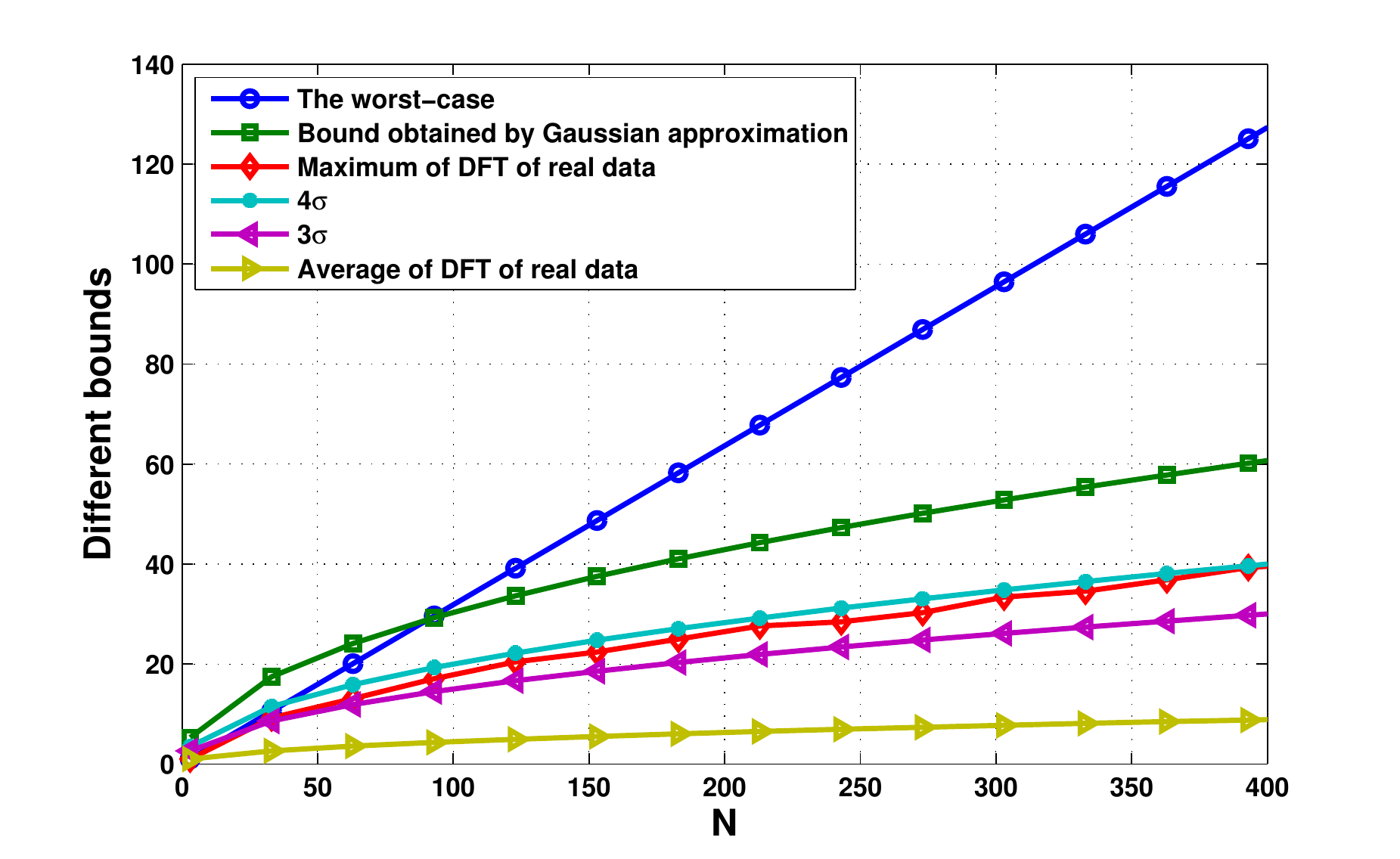}
	\caption{Proposed bounds when the sampling rate is $0.5$.}
	\label{fig5part}
\end{figure}

\begin{figure}[h!]
	\centering
	\includegraphics[width=.91\linewidth]{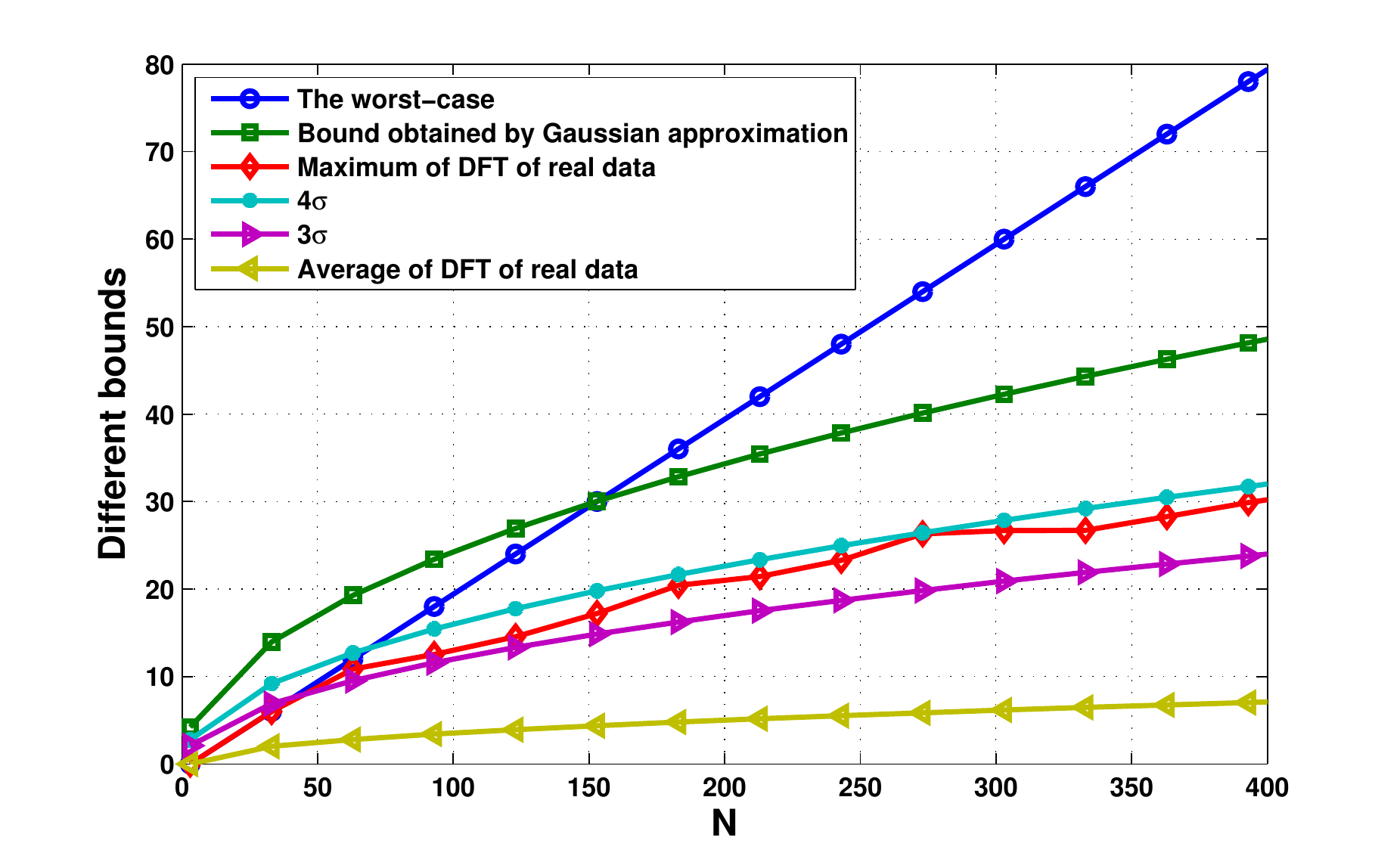}
	\caption{Proposed bounds when the sampling rate is $0.8$.}
	\label{fig8part}
\end{figure}

In this section, we generate random masks with different sizes and calculate maximum and average of their DFT coefficients. Each experiment is repeated $10^5$ times. The maximum and average values along with the proposed bounds are shown in Figs. \ref{fig2part}-\ref{fig8part}. According to the results, the $4-\sigma$ bound follows the maximum value of DFT very well. However, in some cases, it is placed under the maximum curve. The bound obtained by the Gaussian approximation is a confident bound that we are assured that it is placed on top of the maximum value curve. We use $\epsilon=10^{-4}$ for this bound. However, the worst-case bound is useful only for small $N$ because the probability of occurrence of the worst-case is very small for large $N$. For example, when $N=1000$ and the sampling rate is $0.5$, this probability is $2^{-500}$.

\section{Conclusion}\label{sec:conc}
In this paper, we proposed some bounds on the maximum of magnitude of a random mask in Fourier domain. First, we showed that when the worst case occurs. Then, we proposed other bound thanks to approximating the random mask with a Gaussian distribution. Moreover, based on this approximation, the $3-\sigma$ and $4-\sigma$ bounds were also proposed. The numerical examples showed that the bound proposed based on the Gaussian approximation is a confident bound. Moreover, the $4-\sigma$ bound is also a good bound that on average is very close to the reality.

\section*{Acknowledgment}
The author would like to thank Dr. Amin Aminzadeh Gohari for his valuable comments, helps, and suggestions to improve the paper.

\bibliographystyle{IEEEtran}
\bibliography{Citations}

\end{document}